\documentstyle[prl,aps,floats,twocolumn,epsfig]{revtex}
\newcommand{\be}{\begin{equation}}
\newcommand{\ee}{\end{equation}}
\newcommand{\bees}{\begin{eqnarray}}
\newcommand{\ees}{\end{eqnarray}}
\newcommand{\ra}{\rightarrow}

\newcommand{\mpl}{m_{\rm Pl}}
\newcommand{\lpl}{l_{\rm Pl}}

\begin{document}

\par
\begingroup
\twocolumn[%

\begin{flushright}
 IFUP-TH 73/96\\
  December 1996\\
\end{flushright}

{\large\bf\centering\ignorespaces
The graviton background at early times in string cosmology
\vskip2.5pt}
{\dimen0=-\prevdepth \advance\dimen0 by23pt
\nointerlineskip \rm\centering
\vrule height\dimen0 width0pt\relax\ignorespaces
Michele Maggiore
\par}
{\small\it\centering\ignorespaces

INFN, sezione di Pisa, and Dipartimento di Fisica dell'Universit\`{a},\\
piazza Torricelli 2, I-56100 Pisa, Italy.\par}

\par
\bgroup
\leftskip=0.10753\textwidth \rightskip\leftskip
\dimen0=-\prevdepth \advance\dimen0 by17.5pt \nointerlineskip
\small\vrule width 0pt height\dimen0 \relax

We discuss some peculiar properties of the
stochastic graviton background predicted by string cosmology. 
At Planckian times, for the
values of the parameters of the model which are more interesting for
the detection in gravitational wave experiments, the 
 number density of gravitons is parametrically large compared
to Planck density, and is peaked at small energies.
The large parameter is related to the duration of the
string phase and is a characteristic of string cosmology. The
typical interaction time is parametrically larger than the Planck
or string time. Therefore  the shape of the 
graviton spectrum  is not distorted
by thermal effects in the Planck era
and can carry informations on the pre-big bang
phase suggested by string cosmology.

\par\egroup
\vskip2pc]
\thispagestyle{plain}
\endgroup

In recent works, Gasperini and Veneziano have
developed a cosmological model which is based on the low energy
action of string theory~\cite{Ven,GV,review} (see also~\cite{sc}).
Besides its theoretical interest, the model has
 important phenomenological consequences:
the mechanism of amplification of vacuum fluctuations~\cite{Gri,vari}
produces a stochastic graviton background which, for some values of the
parameters of the model, is close to the planned sensitivity
of the LIGO and VIRGO experiments, and within the
sensitivity at which the advanced LIGO detector aims~\cite{BGGV,peak,AB}. 
Electromagnetic  
fluctuations are also amplified, contrarily to what happens in
inflationary cosmology, and might provide the seed for the generation of
galactic magnetic fields~\cite{GGV}.

The model has no initial singularity, but rather
starts from the string perturbative vacuum at $t\ra -\infty$.
Its evolution
at this stage is governed by the string effective action at lowest
order both in the string constant $\alpha '$ and in the loop
corrections, which are governed by $g^2=e^{\phi}$, where $\phi$ is the
dilaton field. The curvature grows until, at a value $\eta =\eta_s$ of
conformal time, it becomes of the order of the string scale. At this
point higher orders in $\alpha '$ in the effective action
become important (while $g^2$ corrections can still be small), 
and the Hubble constant $H$ approaches a constant value
$H_s$~\cite{GMV}. In this phase (``string phase'')
the scale factor $a(\eta )$ and the dilaton $\phi (\eta )$,
in the so called 'string frame', evolve as
\be\label{a}
a(\eta )=-\frac{1}{H_s\eta}\, ,\hspace{10mm}
\phi (\eta)={\rm const.}-2\beta\log\frac{\eta}{\eta_s}\, .
\ee
Here $\beta$ is a dimensionless constant whose value, as the value of
$H_s$, is in principle fixed by the $\alpha '$ corrections. 
This phase lasts up to  some conformal time $\eta =\eta_1$.
The corresponding value of cosmic
time $t_1$ will be of the order of the Planck time $t_{\rm Pl}=\lpl
/c$ or of the string time $\lambda_s/c$, where $\lpl$ is the Planck
length and $\lambda_s$ the string length (we set $\hbar =c =1$ 
in the following). The distinction between $\lambda_s$ and $\lpl$ will
not be important for our discussion. At $\eta =\eta_1$
the coupling $g^2=\exp (\phi )$
becomes large and other effects ($g^2$ corrections, non perturbative
dilaton potential) come into play. 
The mechanism which should mediate
the transition to the standard FRW cosmology with frozen dilaton 
(graceful exit problem~\cite{BV,KMO}) is not yet well understood
although a number of ideas have been suggested recently~\cite{gra}.

The model therefore has two dimensionful parameters $\eta_s,\eta_1$,
with $\eta_s,\eta_1<0$ 
and $|\eta_1|<|\eta_s|$, and a dimensionless parameter 
$\beta$. Denoting by $k$ the comoving
wave number of a graviton and by 
$f$ its frequency red-shifted at the present  epoch,
we can define two new parameters $f_1,f_s$ with dimension of frequency
from $k|\eta_1|=f/f_1, k|\eta_s|=f/f_s$, which we will use in place of
$\eta_1,\eta_s$. Using $2\pi f=k/a(t_{\rm pres})$, where $t_{\rm pres}$
is the present value of cosmic time, and eliminating $\eta_1$ from
eq.~(\ref{a}), we have
\be
k|\eta_1|=\frac{2\pi f}{H_s}\,\frac{a(t_{\rm pres})}{a(t_{1})}\, ,
\ee
which shows that $f_1$ is just  $H_s/(2\pi)$, 
red-shifted from the Planck time $t_1$ to the
 present time. The value of $1/H_s$ is of the order of the
string length $\lambda_s$ times a numerical constant of order one
which is
fixed by the $\alpha '$ corrections~\cite{GMV}, 
and therefore the value of $f_1$ can
be estimated~\cite{BGGV,BMU} to be on the order of 40 GHz. The other
parameter $f_s$ instead can range anywhere between $0<f_s<f_1$. 
From eq.~(\ref{a}), $f_1/f_s=a(\eta_1)/a(\eta_s)$; the value of this
ratio therefore depends on when the string phase begins, $\eta_s$, and
on when it ends, and therefore depends both on the $\alpha '$
corrections and on the mechanism which implements the graceful exit. 
In the absence of a detailed knowledge of these effects, we must
consider $f_1/f_s$ as a free parameter of the model with
$1<f_1/f_s<\infty$. In particular, this parameter can be large, and
has no counterpart in standard cosmology.
The relic graviton spectrum also depends on the constant $\beta$
introduced in eq.~(\ref{a}), or rather on the combination
$\mu =|\beta -(3/2)|$, which takes values in the range $0<\mu
\leq 3/2$~\cite{BGGV}. 
From the phenomenological point of view, an especially interesting
situation is realized when $\mu$ is very close or equal to 3/2 and 
$f_1/f_s$ is very large, at least on the order of $10^{8}$, since in
this case the signal will be close to the planned 
sensitivity of ground based
interferometers as the LIGO and VIRGO experiments. 
If $\mu$ is very close to $3/2$
the experimental bound coming from msec pulsar gives 
the constraint $f_1/f_s < 10^{17}$, otherwise there is 
basically no upper bound to $f_1/f_s$ \cite{BMU}.
As we will see below, the existence of a possible
parametrically large quantity $f_1/f_s$
 has interesting theoretical  consequences.

Recent work has been focused on the energy density of the stocastic
graviton background, which is the quantity of direct experimental
relevance.  Some interesting considerations can however be made
considering instead the number density of gravitons.
The number of gravitons per unit volume at present
time is given by 
\be
N_0=\frac{4\pi}{(2\pi)^3}\int df\, f^2n_f\, ,
\ee
where $n_f$ is the number of gravitons per cell of the phase space and
$f$ is the physical frequency observed today. The number density $N$ at
Planck time $t_1$ is obtained from $N_0$ multiplying it by 
$(a(t_{\rm pres})/a(t_1))^3$. Using the value of $n_f$ computed
in~\cite{BGGV} in the low- and high-frequency limit, and expressing the
result in term of the energy $E$ at Planck time, 
$E=2\pi f a(t_{\rm pres})/a(t_1)$, we obtain
\be\label{dN}
\frac{dN}{d\log E}\simeq \frac{1}{\lambda_s^3}\times 
\left\{ 
\begin{array}{ll}
c_1\left(\frac{f_1}{f_s}\right)^{2\mu}
\left(\frac{E}{H_s}\log\frac{E}{H_s}\right)^2
\hspace*{5mm}\frac{E}{H_s} \ll \frac{f_s}{f_1}\\ 
c_2\left(\frac{E}{H_s}\right)^{2-2\mu}\,
\hspace*{11mm} \frac{f_s}{f_1}\ll \frac{E}{H_s}<1\, .
\end{array}
\right.
\ee
The spectrum has a cutoff at $E\simeq H_s$.
The numerical constants $c_1,c_2$ and the behavior in
the intermediate frequency region can  be obtained 
from ref.~\cite{BMU},
\bees
c_1(\mu )&=&\frac{(2\mu -1)^2 (2\mu\alpha
-1+\alpha)^2}{256\pi^3\mu^2\alpha}, \hspace{3mm}\alpha =1/(1+\sqrt{3})
\nonumber\\
c_2(\mu )&=&\frac{1}{64\pi^3}\, 2^{2\mu} (2\mu -1)^2\Gamma^2(\mu )\, .
\ees
For  values of $\mu$ not very close to zero
$c_1,c_2$ are small numbers: for
instance $c_2(1.5)\simeq 0.012$ and $c_2(1)\simeq 0.002$. Similarly
$c_1(\mu )$ has typical values $O(10^{-3})$.

From eq.~(\ref{dN}) we see that the spectrum of 
graviton number density at
the Planckian era, per unit logarithmic interval of energy, has a
completely different behavior depending on whether $0<\mu \leq 1$ or
$1<\mu \leq 3/2$. In the former case it is a monotonically growing function
of the energy up to the cutoff energy $E\sim H_s$, and its value at
 the maximum is of order $c_2/\lambda_s^3$.
In the latter case it has a
maximum at $E\sim (f_s/f_1) H_s$ and then decreases. Fig.~1 shows the
behavior of $\lambda_s^3dN/d\log E$
for $\mu =1.5$ and $\mu =0.75$ for a moderate value of
$f_1/f_s =10^2$. Note that the result for $\mu =0.75$ has been
multiplied by $10^3$ in order to plot it on the same scale as 
$\mu =1.5$.  We have used the analytical results of ref.~\cite{BMU}
 for the energy dependence in the whole region.

\begin{figure}[t]
\epsfxsize=3in
\centerline{\psfig{figure=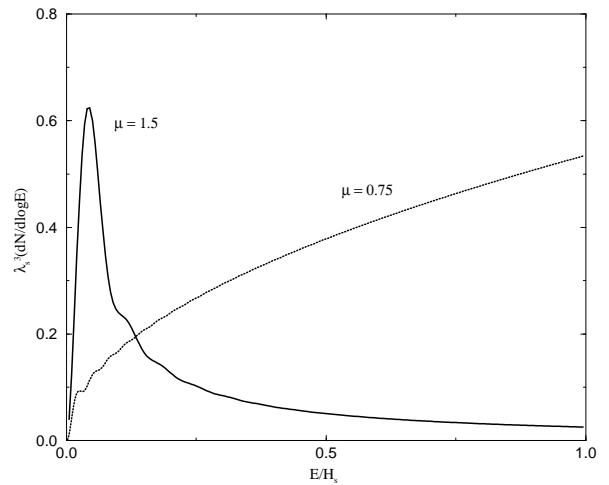,width=3in,angle=-90}}
\caption{\sl $\lambda_s^3dN/d\log E$
for $\mu =1.5$ and $\mu =0.75$,  for 
$f_1/f_s =10^2$. The data for $\mu =0.75$ have been
multiplied by $10^3$ in order to plot them on the same scale as
$\mu =1.5$. For this value of $f_1/f_s$ the height of the
peak is still smaller than one. For $f_1/f_s =10^8$ the value at the
peak becomes of order $10^6$.
}
\end{figure}

The value at the peak for $\mu >1$ is 
\be
\left(\frac{dN}{d\log E}\right)_{\rm max}\sim\frac{c_2}{\lambda_s^3}
\left( \frac{f_1}{f_s}\right)^{2\mu -2}
\ee
and is much larger than $1/\lambda_s^3$ if $f_1\gg f_s$. The
integrated density of gravitons for $\mu$ generic is
\be\label{7}
N=\int d\log E\, \frac{dN}{d\log E}\sim \frac{1}{\lambda_s^3}
\left[ c_1+c_2
\left(\frac{f_1}{f_s}\right)^{2\mu -2}\right]\, .
\ee
We see that if $\mu <1$ we have about $10^{-3}$
graviton in a string volume $\lambda_s^3$, while if 
$\mu>1$ and $f_1/f_s$ is sufficently
large, we have a parametrically large density.

It is interesting to ask whether graviton-graviton interactions at the
Planckian era are strong enough to establish thermal equilibrium. The
condition for thermalization is that the typical interaction time
$\tau$ be much smaller than the expansion rate $1/H_s$. The average
interaction time for a graviton of energy $E$ can be estimated from
\be
\frac{1}{\tau}=\int d\log E'\, \sigma (E,E')\frac{dN(E')}{d\log E'}\, .
\ee
Here $\sigma (E,E')$ is the  cross section for scattering of 
two gravitons with energies $E,E'$, and is of the order of
 $\sigma (E,E')\sim G^2s\sim\lpl^2(EE'/\mpl^2)$, where 
$\mpl$ the Planck mass,
$G$ the gravitational constant and $s$ the square of
the c.m. energy. Using eq.~(\ref{dN})
we find separately the contribution to the
integral from the scattering on soft and on hard gravitons, 
\be
\frac{1}{\tau_{\rm soft}}\sim E\left(\frac{f_1}{f_s}\right)^{2\mu
-3}\, ,
\hspace{5mm}
\frac{1}{\tau_{\rm hard}}\sim E
\ee
so that for $f_1/f_s$ large
the main contribution comes from scattering on hard gravitons and
 $1/\tau\sim E$. In the case $\mu <1$ most gravitons have
a  Planckian energy and therefore the typical value of $\tau$ is of
order one in Planck or string units, and it is therefore of the order
of the expansion time. This is the typical situation which one
meets in standard cosmology, extrapolating the low energy results
close to the Planck era~\cite{NZ}: 
there is no parametrically large number and
therefore all quantities
are of order one in Planck units. It is therefore impossible to draw
definite conclusions, since the establishing of thermal equilibrium
depends on the balance between numbers which we are unable to compute.
Of course, one might observe that the overall  factors
$c_1(\mu ),c_2(\mu )$  are numerically small and consider this 
as an indication that the
gravitons do not thermalize, but such a conclusion would not be 
reliable.

For $\mu >1$ and $f_1/f_s$ very large, the situation is
different and much more interesting. Almost all
gravitons have a  small energy, $E\sim f_s/f_1$ in Planck
units, and therefore the typical interaction time 
$\tau\sim E^{-1}\sim (f_1/f_s)\lpl$
is parametrically larger than $1/H_s\sim\lambda_s$ 
(note also that the use of the low energy cross section is
more justified in this case).
We can therefore draw the conclusion that thermal equilibrium
is not reached, and graviton-graviton interactions at the Planck era
are neglegible. This has an important phenomenological consequence. It
implies that the relic graviton spectrum is not distorted by
interactions during the Planck epoch, and therefore 
graviton-graviton interactions do not mask the signal coming from the
pre-big bang era.

We conclude stressing that at the Planckian epoch
the relic graviton background for $1<\mu\leq 3/2$
and $f_1/f_s\gg 1$ is in a rather remarkable state. Its 
number density is
huge, being parametrically larger than one in Planck units. 
However, it is made of parametrically soft gravitons, and the
competition between these two effects results in a  very weakly
interacting system. 
The typical de Broglie wavelength associated with
thermal motion is $\lambda_T\sim 1/E\gg \lpl$ while the average spacing
between gravitons is $d\sim N^{-1/3}\ll \lpl$, and
therefore $\lambda_T\gg d$; this suggests that this
system is appropriately described by a collective wave-function
describing the state of the gravitational field rather than in terms
of single gravitons. Coherent quantum effects  are therefore probably
important at this stage.
In the formal limit $f_1/f_s\ra \infty$ the total number of
gravitons diverges, see eq.~(\ref{7}), and almost 
all of them accumulate in the peak at the energy $E_{\rm peak}
\sim (f_s/f_1)H_s\ra 0$.
This fact, togheter with the condition $\lambda_T\gg d$, shows 
an interesting analogy with
Bose-Einstein condensation. The analogy is however only partial,
because we are not dealing with thermal distributions.

\vspace*{5mm}

I thank Gabriele Veneziano for very useful discussions.

\end{document}